\documentclass[draft,tightenlines,nofootinbib,preprint,aps,eqsecnum,amsmath,amssymb]{revtex4}

\newcommand{\beq}{\begin{equation}}
\newcommand{\eeq}{\end{equation}}
\newcommand{\bea}{\begin{eqnarray}}
\newcommand{\eea}{\end{eqnarray}}

\newcommand{\sgn}{\epsilon}
\newcommand{\eo}{{}^4{\buildrel \circ \over E}}

\begin{document}

\title{Non-Inertial Frames in Special and General Relativity}

\medskip

\author{Luca Lusanna}

\affiliation{ Sezione INFN di Firenze\\ Polo Scientifico\\ Via Sansone 1\\
50019 Sesto Fiorentino (FI), Italy\\ Phone: 0039-055-4572334\\
FAX: 0039-055-4572364\\ E-mail: lusanna@fi.infn.it}

\today

\begin{abstract}

There is a review of what is known about global non-inertial frames
in special and general relativity.
\bigskip

Proceedings of the SIGRAV School "Gravity: Where Do We Stand?", Villa Olmo (Como) 2009,
eds. V.Gorini, U.Moschella and R.Peron, to appear in Lecture Notes Physics, Springer.

\end{abstract}

\maketitle

\vfill\eject

\section{Introduction}

The aim of this contribution is to clarify what is known about
non-inertial frames in special (SR) and general (GR) relativity.
This topic is rarely discussed and till recently there was no
attempt to develop a consistent general theory. All the results of
the standard model of elementary particles are defined in the
inertial frames of Minkowski space-time. Only at the level of
neutron, atomic and molecular physics one needs a local study of
non-inertial frames in SR, for instance the rotating ones for the
Sagnac effect.

\medskip

Moreover, relativistic metrology \cite{1} and space physics around
the Earth and in the Solar System \cite{2} must take into account
the gravitational field and the Post-Newtonian (PN) limit of GR, a
theory in which global inertial frames are forbidden by the
equivalence principle. In Einstein GR the gauge group of its
Lagrangian formulation, the diffeomorphism group, implies that the
4-coordinates of the space-time (and therefore the local
non-inertial frames) are {\it gauge variables}. As a consequence,
one would like to describe the effects of the physical degrees of
freedom of the gravitational field by means of 4-scalars. This is an
open theoretical problem. The praxis of experimentalists, who do not
know which is the correct formulation of GR among the existing ones,
is completely different.

\medskip

Inside the Solar System the experimental localization of macroscopic
classical objects is unavoidably done by choosing some {\it
convention} for the local 4-coordinates of space-time. Atomic
physicists, NASA engineers and astronomers have chosen a series of
reference frames and standards of time and length suitable for the
existing technology \cite{1,2}. These conventions determine certain
Post-Minkowskian (PM) 4-coordinate systems of an asymptotically
Minkowskian space-time, in which the instantaneous 3-spaces are not
strictly Euclidean. Then these reference frames are seen as a local
approximation of a Celestial Reference Frame (ICRS), where however
the space-time has become a cosmological Friedman-Robertson-Walker
(FRW) one, which is only conformally asymptotically Minkowskian at
spatial infinity. A search of a consistent patching of the
4-coordinates from inside the Solar System to the rest of the
universe will start when the data from the future GAIA mission
\cite{3} for the cartography of the Milky Way will be available.
This will allow a PM definition of a Galactic Reference System
containing at leat our galaxy. Let us remark that notwithstanding
the FRW instantaneous 3-spaces are not strictly Euclidean, all the
books on galaxy dynamics describe the galaxies by means of Kepler
theory in Galilei space-time.

\medskip

A well posed formulation of a PM ICRS (a global non-inertial frame
for the 3-universe) would also be needed to face the main open
problem of astrophysics, namely the dominance of {\it dark}
entities, the dark matter and the dark energy, in the existing
description of the universe given by the standard $\Lambda$CDM
cosmological model \cite{4} based on the cosmological principle
(homogeneity and isotropy of the space-time), which selects the
class of Friedmann-Robertson-Walker (FWR) space-times. After the
transition from quantum cosmology to classical astrophysics, with
the Heisenberg cut roughly located  at a suitable cosmic time
($\approx 10^5$ years after the big bang) and at the recombination
surface identified by the cosmic microwave background (CMB), one has
a description of the universe in which the known forms of baryonic
matter and radiation contribute only with a few percents of the
global budget. One has a great variety of models trying to explain
the composition of the universe in accelerated expansion (based on
data on high red-shift supernovae, galaxy clusters and CMB): WIMPS
(mainly super-symmetric particles), $f(R)$ modifications of Einstein
gravity (with a modified Newton potential), MOND (with a
modification of Newton law),... for dark matter; cosmological
constant, string theory, back-reaction (spatial averages,
non-linearity of Einstein equations), inhomogeneous space-times
(Lemaitre-Tolman-Bondi, Szekeres), scalar fields (quintessence,
k-essence, phantom), fluids (Chaplygin fluid), .... for dark energy.

\medskip

A PM ICRS would allow to interpret the astronomical data
(luminosity, light spectrum, angles) on the 2-dimensional sky vault
in a more realistic way (taking into account the inhomogeneities in
the 3-universe) than in the nearly flat 3-spaces (as required by CMB
data) of FWR space-times. In particular one needs new standards of
time and length like the cosmic time and the luminosity distance
extending the standard relativistic metrology inside the Solar
System.

\medskip

All these open problems justify the following description of what is
known about non-inertial frames in SR and GR.

\section{Non-Inertial Frames in Special Relativity}

In non-relativistic (NR) Newtonian physics isolated systems are
described in Galilei space-time, where both time and the
instantaneous Euclidean 3-spaces are absolute quantities. As a
consequence, the transition from the description of the system in NR
inertial frames to its description in rigid non-inertial frames can
be done by defining the following 3-coordinate transformation

\beq
 x^i = y^i(t) + \sigma^r\, R_{ri}(t).
 \label{1.1}
 \eeq

\noindent Here $x^i$'s are inertial Cartesian 3-coordinates centered
on an inertial observer, while $\sigma^r$'s are rigid non-inertial
3-coordinates centered on an arbitrary observer whose trajectory is
described by the Cartesian 3-coordinates $y^i(t)$ in the inertial
frame. This accelerated observer has a 3-velocity, which can be
conveniently written in the form $v^i(t) = R_{ij}(t)\, {{d
y^j(t)}\over {dt}}$. $R(t)$ is a time-dependent rotation matrix
($R^{-1} = R^T$), which can be parametrized with three
time-dependent Euler angles. The angular velocity of the rotating
frame is $\omega^i(t) = {1\over 2}\, \epsilon^{ijk}\,
\Omega_{jk}(t)$ with $\Omega_{jk}(t) = - \Omega_{kj}(t) = \Big({{d
R(t)}\over {dt}}\, R^T(t)\Big)_{jk}$.

\bigskip

A particle of mass $m_o$ with inertial Cartesian 3-coordinates
$x^i_o(t)$ is described in the non-inertial frame by 3-coordinates
$\eta^r(t)$ such that

\beq
 x^i_o(t) = y^i(t) + \eta^r(t)\, R_{ri}(t).
 \label{1.2}
 \eeq

As shown in every book on Newtonian mechanics a particle satisfying
the equation of motion $m_o\, {{d^2\, {\vec x}_o(t)}\over {dt^2}} =
- {{\partial\, V(t, x^j_o(t))}\over {\partial\, {\vec x}_o}}$, if an
external potential $V(t, x^k_o(t)) = \tilde V(t, \eta^r(t))$ is
present, will satisfy the following equation of motion in the rigid
non-inertial frame

\bea
 m_o\, {{d^2\, \vec \eta(t)}\over {dt^2}} &=&
 - {{\partial\, \tilde V(t, \eta^r(t))}\over {\partial\, {\vec \eta}_o}}
 -\nonumber \\
 &-& m_o\, \Big[{{d \vec v(t)}\over {dt}} + \vec \omega(t) \times \vec
 v(t) + {{d \vec \omega(t)}\over {dt}} \times \vec \eta(t) +\nonumber \\
 &+& 2\, \vec \omega(t) \times {{d \vec \eta(t)}\over {dt}} +
 \vec \omega(t) \times [\vec \omega(t) \times \vec \eta(t) \Big],
 \label{1.3}
 \eea

\noindent where the standard Euler, Jacobi, Coriolis and centrifugal
{\it inertial forces} associated with the linear acceleration of the
non-inertial observer and with the angular velocity of the rotating
frame are present.\bigskip

In Ref.\cite{5} there is the extension to non-rigid non-inertial
frames in which Eq.(\ref{1.1}) is replaced by $x^i = {\cal A}^i(t,
\sigma^r)$ with ${\cal A}^i$ arbitrary functions well behaved at
spatial infinity. For instance, a differentially rotating
non-inertial frame is described by Eq.(\ref{1.1}) with a
point-dependent rotation matrix $R(t, \sigma^r)$.
\bigskip

To go from NR inertial frames to the non-rigid non-inertial ones one
has to replace the group of Galilei transformations, connecting the
NR inertial frames, with some subgroup of the group of
3-diffeomorphisms of the Euclidean 3-space.

\bigskip

The transition to SR is highly non trivial because, due to the
Lorentz signature of Minkowski space-time, time is no more absolute
and there is no notion of instantaneous 3-space: the only intrinsic
structure is the conformal one, i.e. the light-cone as the locus of
incoming and outgoing radiation. {\it A convention on the
synchronization of clocks is needed to define an instantaneous
3-space}. For instance the {\it 1-way velocity of light} from one
observer A to an observer B has a meaning only after a choice of a
convention for synchronizing the clock in A with the one in B.
Therefore the crucial quantity in SR is the {\it 2-way (or round
trip) velocity of light $c$} involving only one clock. It is this
velocity  which is isotropic and constant in SR and replaces the
standard of length in relativistic metrology \footnote{See
Ref.\cite{1} for an updated review on relativistic metrology on
Earth and in the Solar System.}.
\medskip

Einstein convention for the synchronization of clocks in Minkowski
space-time uses the 2-way velocity of light to identify the
Euclidean 3-spaces of the inertial frames centered on an inertial
observer A by means of only its clock. The inertial observer A sends
a ray of light at $x^o_i$ towards the (in general accelerated)
observer B; the ray is reflected towards A at a point P of B
world-line and then reabsorbed by A at $x^o_f$; by convention P is
synchronous with the mid-point between emission and absorption on
A's world-line, i.e. $x^o_P = x^o_i + {1\over 2}\, (x^o_f - x^o_i) =
{1\over 2}\, (x^o_i + x^o_f)$. This convention selects the Euclidean
instantaneous 3-spaces $x^o = ct = const.$ of the inertial frames
centered on A. Only in this case the one-way velocity of light
between A and B coincides with the two-way one, $c$. However if the
observer A is accelerated, the convention can breaks down due the
possible appearance of coordinate singularities.

\bigskip

The existing coordinatizations, like either Fermi or Riemann-normal
coordinates, hold only locally They are based on the {\it 1+3 point
of view}, in which only the world-line of a time-like observer is
given. In each point of the world-line the observer 4-velocity
determines an orthogonal 3-dimensional space-like tangent
hyper-plane, which is identified with an instantaneous 3-space.
However, these tangent planes intersect at a certain distance from
the world-line (the so-called acceleration length depending upon the
4-acceleration of the observer \cite{6}), where 4-coordinates of the
Fermi type develop a coordinate singularity. Another type of
coordinate singularity is developed in rigidly rotating coordinate
systems at a distance $r$ from the rotation axis where $\omega\, r =
c$ ($\omega$ is the angular velocity and $c$ the two-way velocity of
light). This is the so-called "horizon problem of the rotating
disk": a time-like 4-velocity becomes a null vector at $\omega\, r =
c$, like it happens on the horizon of a black-hole. See Ref.\cite{7}
for a classification of the possible pathologies of non-inertial
frames and on how to avoid them.

\medskip

As a consequence, a theory of global non-inertial frames in
Minkowski space-time has to be developed in a metrology-oriented way
to overcame the pathologies of the 1+3 point of view. This has been
done in the papers of Ref.\cite{7} by using the {\it 3+1 point of
view} in which, besides the world-line of a time-like observer, one
gives a global nice foliation of the space-time with instantaneous
3-spaces.

\medskip

Assume that the world-line $x^{\mu}(\tau)$ of an arbitrary time-like
observer carrying a standard atomic clock is given: $\tau$ is an
arbitrary monotonically increasing function of the proper time of
this clock. Then one gives an admissible 3+1 splitting of Minkowski
space-time, namely a nice foliation with space-like instantaneous
3-spaces $\Sigma_{\tau}$. It is the mathematical idealization of a
protocol for clock synchronization: all the clocks in the points of
$\Sigma_{\tau}$ sign the same time of the atomic clock of the
observer \footnote{It is the non-factual idealization required by
the Cauchy problem generalizing the existing protocols for building
coordinate system inside the future light-cone of a time-like
observer.}. On each 3-space $\Sigma_{\tau}$ one chooses curvilinear
3-coordinates $\sigma^r$ having the observer as origin. These are
the Lorentz-scalar and observer-dependent {\it radar 4-coordinates}
$\sigma^A = (\tau; \sigma^r)$, first introduced by Bondi \cite{8}.

\bigskip

If $x^{\mu} \mapsto \sigma^A(x)$ is the coordinate transformation
from the Cartesian 4-coordinates $x^{\mu}$ of a reference inertial
observer to radar coordinates, its inverse $\sigma^A \mapsto x^{\mu}
= z^{\mu}(\tau ,\sigma^r)$ defines the {\it embedding} functions
$z^{\mu}(\tau ,\sigma^r)$ describing the 3-spaces $\Sigma_{\tau}$ as
embedded 3-manifold into Minkowski space-time. The induced 4-metric
on $\Sigma_{\tau}$ is the following functional of the embedding
${}^4g_{AB}(\tau ,\sigma^r) = [z^{\mu}_A\, \eta_{\mu\nu}\,
z^{\nu}_B](\tau ,\sigma^r)$, where $z^{\mu}_A = \partial\,
z^{\mu}/\partial\, \sigma^A$ and ${}^4\eta_{\mu\nu} = \sgn\, (+---)$
is the flat metric \footnote{$\sgn = \pm 1$ according to either the
particle physics $\sgn = 1$ or the general relativity $\sgn = - 1$
convention.}. While the 4-vectors $z^{\mu}_r(\tau ,\sigma^u)$ are
tangent to $\Sigma_{\tau}$, so that the unit normal $l^{\mu}(\tau
,\sigma^u)$ is proportional to $\epsilon^{\mu}{}_{\alpha
\beta\gamma}\, [z^{\alpha}_1\, z^{\beta}_2\, z^{\gamma}_3](\tau
,\sigma^u)$, one has $z^{\mu}_{\tau}(\tau ,\sigma^r) = [N\, l^{\mu}
+ N^r\, z^{\mu}_r](\tau ,\sigma^r)$  with $N(\tau ,\sigma^r) =
\sgn\, [z^{\mu}_{\tau}\, l_{\mu}](\tau ,\sigma^r) = 1 + n(\tau,
\sigma^r)$ and $N_r(\tau ,\sigma^r) = - \sgn\, g_{\tau r}(\tau
,\sigma^r)$ being the lapse and shift functions
respectively.\bigskip

As a consequence, the components of the 4-metric ${}^4g_{AB}(\tau
,\sigma^r )$ have the following expression

\bea
 \sgn\, {}^4g_{\tau\tau} &=& N^2 - N_r\, N^r,\qquad
  - \sgn\, {}^4g_{\tau r} = N_r = {}^3g_{rs}\, N^s,\nonumber \\
 {}^3g_{rs} &=& - \sgn\, {}^4g_{rs} = \sum_{a=1}^3\, {}^3e_{(a)r}\,
 {}^3e_{(a)s} = \nonumber \\
 &=& {\tilde \phi}^{2/3}\, \sum_{a=1}^3\,
 e^{2\, \sum_{\bar b =1}^2\, \gamma_{\bar ba}\, R_{\bar b}}\,
 V_{ra}(\theta^i)\, V_{sa}(\theta^i),
 \label{1.4}
 \eea

\noindent where ${}^3e_{(a)r}(\tau ,\sigma^u)$ are cotriads on
$\Sigma_{\tau}$, ${\tilde \phi}^2(\tau ,\sigma^r) = det\,
{}^3g_{rs}(\tau ,\sigma^r)$ is the 3-volume element on
$\Sigma_{\tau}$, $\lambda_a(\tau ,\sigma^r) = [{\tilde \phi}^{1/3}\,
e^{\sum_{\bar b =1}^2\, \gamma_{\bar ba}\, R_{\bar b}}](\tau
,\sigma^r)$ are the positive eigenvalues of the 3-metric
($\gamma_{\bar aa}$ are suitable numerical constants) and
$V(\theta^i(\tau ,\sigma^r))$ are diagonalizing rotation matrices
depending on three Euler angles.\medskip

Therefore starting from the {\it four} independent embedding
functions $z^{\mu}(\tau, \sigma^r)$ one obtains the {\it ten}
components ${}^4g_{AB}$ of the 4-metric (or the quantities $N$,
$N_r$, $\tilde \phi$, $R_{\bar a}$, $\theta^i$), which play the role
of the {\it inertial potentials} generating the relativistic
apparent forces in the non-inertial frame. It can be shown \cite{7}
that the usual NR Newtonian inertial potentials are hidden in the
functions $N$, $N_r$ and $\theta^i$. The extrinsic curvature tensor
${}^3K_{rs}(\tau, \sigma^u) = [{1\over {2\, N}}\, (N_{r|s} + N_{s|r}
- \partial_{\tau}\, {}^3g_{rs})](\tau, \sigma^u)$, describing the
{\it shape} of the instantaneous 3-spaces of the non-inertial frame
as embedded 3-sub-manifolds of Minkowski space-time, is a secondary
inertial potential, functional of the ten inertial potentials
${}^4g_{AB}$. Now a relativistic positive-energy scalar particle
with world-line $x^{\mu}_o(\tau)$ is  described by 3-coordinates
$\eta^r(\tau)$ defined by $x^{\mu}_o(\tau) = z^{\mu}(\tau,
\eta^r(\tau))$, satisfying  equations of motion containing
relativistic inertial forces whose non-relativistic limit reproduces
Eq.(\ref{1.3}) as shown in Ref.\cite{5,7}.

\bigskip

The foliation is nice and admissible if it satisfies the conditions:
\hfill\medskip

1) $N(\tau ,\sigma^r) > 0$ in every point of $\Sigma_{\tau}$ so that
the 3-spaces never intersect, avoiding the coordinate singularity of
Fermi coordinates;\hfill\medskip

2) $\sgn\, {}^4g_{\tau\tau}(\tau ,\sigma^r) > 0$, so to avoid the
coordinate singularity of the rotating disk, and with the
positive-definite 3-metric ${}^3g_{rs}(\tau ,\sigma^u) = - \sgn\,
{}^4g_{rs}(\tau ,\sigma^u)$ having three positive eigenvalues (these
are the M$\o$ller conditions \cite{9});\hfill\medskip

3) all the 3-spaces $\Sigma_{\tau}$ must tend to the same space-like
hyper-plane at spatial infinity with a unit  normal
$\epsilon^{\mu}_{\tau}$, which is the time-like 4-vector of a set of
asymptotic ortho-normal tetrads $\epsilon^{\mu}_A$. These tetrads
are carried by asymptotic inertial observers and the spatial axes
$\epsilon^{\mu}_r$ are identified by the fixed stars of star
catalogues. At spatial infinity the lapse function tends to $1$ and
the shift functions vanish.

\bigskip

By using the asymptotic tetrads $\epsilon^{\mu}_A$ one can give the
following parametrization of the embedding functions

\bea
 z^{\mu}(\tau, \sigma^r) &=& x^{\mu}(\tau) + \epsilon^{\mu}_A\,
 F^A(\tau, \sigma^r),\qquad  F^A(\tau, 0) = 0,\nonumber \\
 &&{}\nonumber \\
 x^{\mu}(\tau) &=& x^{\mu}_o + \epsilon^{\mu}_A\, f^A(\tau),
 \label{1.5}
 \eea

\noindent where $x^{\mu}(\tau)$ is the world-line of the observer.
The functions $f^A(\tau)$ determine the 4-velocity $u^{\mu}(\tau) =
{\dot x}^{\mu}(\tau)/ \sqrt{\sgn\, {\dot x}^2(\tau)}$ (${\dot
x}^{\mu}(\tau) = {{d x^{\mu}(\tau)}\over {d\tau}}$) and the
4-acceleration $a^{\mu}(\tau) = {{d u^{\mu}(\tau)}\over {d\tau}}$ of
the observer.
\bigskip

The M$\o$ller conditions are non-linear differential conditions on
the functions $f^A(\tau)$ and $F^A(\tau, \sigma^r)$, so that it is
very difficult to construct explicit examples of admissible 3+1
splittings. When these conditions are satisfied Eqs.(\ref{1.5})
describe a global non-inertial frame in Minkowski space-time.

\bigskip

Till now the solution of M$\o$ller conditions is known in the
following two cases in which the instantaneous 3-spaces are parallel
Euclidean space-like hyper-planes not equally spaced due to a linear
acceleration.\medskip

A)  {\it Rigid non-inertial reference frames with translational
acceleration}. An example are the following embeddings

\medskip

\bea
 z^{\mu}(\tau ,\sigma^u ) &=& x^{\mu}_o +
\epsilon^{\mu}_{\tau}\, f(\tau ) + \epsilon^{\mu}_r\,
\sigma^r,\nonumber \\
 &&{}\nonumber \\
 &&g_{\tau\tau}(\tau ,\sigma^u ) = \sgn\,
 \Big({{d f(\tau )}\over {d\tau}}\Big)^2,\quad g_{\tau r}(\tau ,\sigma^u )
 =0,\quad g_{rs}(\tau ,\sigma^u ) = -\sgn\, \delta_{rs}.\nonumber \\
 &&{}
 \label{1.6}
 \eea

\medskip

This is a foliation with parallel hyper-planes with normal $l^{\mu}
= \epsilon^{\mu}_{\tau} = const.$ and with the time-like observer
$x^{\mu}(\tau ) = x^{\mu}_o + \epsilon^{\mu}_{\tau}\, f(\tau )$ as
origin of the 3-coordinates. The hyper-planes have translational
acceleration ${\ddot x}^{\mu}(\tau ) = \epsilon^{\mu}_{\tau}\, \ddot
f(\tau )$, so that they are not uniformly distributed like in the
inertial case $f(\tau ) = \tau$.

\bigskip

B) {\it Differentially rotating non-inertial frames} without the
coordinate singularity of the rotating disk.  The embedding defining
this frames is

\begin{eqnarray*}
 z^{\mu}(\tau ,\sigma^u ) &=& x^{\mu}(\tau ) + \epsilon^{\mu}_r\,
R^r{}_s(\tau , \sigma )\, \sigma^s \, \rightarrow_{\sigma
\rightarrow \infty}\,
x^{\mu}(\tau) + \epsilon^{\mu}_r\, \sigma^r,\nonumber \\
 &&{}\nonumber \\
 R^r{}_s(\tau ,\sigma ) &=& R^r{}_s(\alpha_i(\tau,\sigma )) =
 R^r{}_s(F(\sigma )\, {\tilde \alpha}_i(\tau)),\nonumber \\
 &&{}\nonumber \\
 &&0 < F(\sigma ) < {1\over {A\, \sigma}},\qquad {{d\, F(\sigma
 )}\over {d\sigma}} \not= 0\,\, (Moller\,\, conditions),
 \end{eqnarray*}

\bea
 z^{\mu}_{\tau}(\tau ,\sigma^u) &=& {\dot x}^{\mu}(\tau ) -
 \epsilon^{\mu}_r\,  R^r{}_s(\tau
 ,\sigma )\, \delta^{sw}\, \epsilon_{wuv}\, \sigma^u\, {{\Omega^v(\tau
 ,\sigma )}\over c},\nonumber \\
  z^{\mu}_r(\tau ,\sigma^u) &=& \epsilon^{\mu}_k\, R^k{}_v(\tau
 ,\sigma )\, \Big(\delta^v_r + \Omega^v_{(r) u}(\tau ,\sigma )\,
 \sigma^u\Big),
 \label{1.7}
 \eea

\noindent where $\sigma = |\vec \sigma |$ and $R^r{}_s(\alpha_i(\tau
,\sigma ))$ is a rotation matrix satisfying the asymptotic
conditions $R^r{}_s(\tau , \sigma)\, {\rightarrow}_{\sigma
\rightarrow \infty} \delta^r_s$, $\partial_A\, R^r{}_s(\tau ,\sigma
)\, {\rightarrow}_{\sigma \rightarrow \infty}\, 0$, whose Euler
angles have the expression $\alpha_i(\tau ,\vec \sigma ) = F(\sigma
)\, {\tilde \alpha}_i(\tau )$, $i=1,2,3$. The unit normal is
$l^{\mu} = \epsilon^{\mu}_{\tau} = const.$ and the lapse function is
$1 + n(\tau ,\sigma^u) = \sgn\, \Big(z^{\mu}_{\tau}\,
l_{\mu}\Big)(\tau ,\sigma^u) = \sgn\, \epsilon^{\mu}_{\tau}\, {\dot
x}_{\mu}(\tau ) > 0$. In Eq.(\ref{1.7}) one uses the notations
$\Omega_{(r)}(\tau ,\sigma ) = R^{-1}(\tau ,\vec \sigma )\,
\partial_r\, R(\tau ,\sigma )$ and $\Big(R^{-1}(\tau ,\sigma )\,
\partial_{\tau}\, R(\tau ,\sigma )\Big)^u{}_v = \delta^{um}\,
\epsilon_{mvr}\, {{\Omega^r(\tau ,\sigma)}\over c}$, with
$\Omega^r(\tau ,\sigma ) = F(\sigma )\, \tilde \Omega (\tau ,\sigma
)$ ${\hat n}^r(\tau ,\sigma )$ \footnote{${\hat n}^r(\tau ,\sigma )$
defines the instantaneous rotation axis and $0 < \tilde \Omega (\tau
,\sigma ) < 2\, max\, \Big({\dot {\tilde \alpha}}(\tau ), {\dot
{\tilde \beta}}(\tau ), {\dot {\tilde \gamma}}(\tau )\Big)$.} being
the angular velocity. The angular velocity vanishes at spatial
infinity and has an upper bound proportional to the minimum of the
linear velocity $v_l(\tau ) = {\dot x}_{\mu}\, l^{\mu}$ orthogonal
to the space-like hyper-planes. When the rotation axis is fixed and
$\tilde \Omega (\tau ,\sigma ) = \omega = const.$, a simple choice
for the function $F(\sigma )$ is $F(\sigma ) = {1\over {1 +
{{\omega^2\, \sigma^2}\over {c^2}}}}$ \footnote{Nearly rigid
rotating systems, like a rotating disk of radius $\sigma_o$, can be
described by using a function $F(\sigma )$ approximating the step
function $\theta (\sigma - \sigma_o)$.}.\medskip

To evaluate the non-relativistic limit for $c \rightarrow \infty$,
where $\tau = c\, t$ with $t$ the absolute Newtonian time, one
chooses the gauge function $F(\sigma ) = {1\over {1 + {{\omega^2\,
\sigma^2}\over {c^2}}}}\, \rightarrow_{c \rightarrow \infty}\, 1 -
{{ \omega^2\, \sigma^2}\over {c^2}} + O(c^{-4})$. This implies that
the corrections to rigidly-rotating non-inertial frames coming from
M$\o$ller conditions are of order $O(c^{-2})$ and become important
at the distance from the rotation axis where the horizon problem for
rigid rotations appears.

\bigskip

As shown in the first paper in Refs.\cite{7},  {\it global rigid
rotations are forbidden in relativistic theories}, because, if one
uses the embedding $z^{\mu}(\tau ,\sigma^u)= x^{\mu}(\tau ) +
\epsilon^{\mu}_r\, R^r{}_s(\tau )\, \sigma^s$ describing a global
rigid rotation with angular velocity $\Omega^r = \Omega^r(\tau )$,
then the resulting $g_{\tau\tau}(\tau ,\sigma^u)$ violates M$\o$ller
conditions, because it vanishes at $\sigma = \sigma_R = {1\over
{\Omega (\tau )}}\, \Big[\sqrt{{\dot x}^2(\tau ) + [{\dot
x}_{\mu}(\tau )\, \epsilon^{\mu}_r\, R^r{}_s(\tau )\, (\hat \sigma
\times \hat \Omega (\tau ))^r]^2}$ $- {\dot x}_{\mu}(\tau )\,
\epsilon^{\mu}_r\, R^r{}_s(\tau )\, (\hat \sigma \times \hat \Omega
(\tau ))^r \Big]$ ( $\sigma^u = \sigma\, {\hat \sigma}^u$, $\Omega^r
= \Omega\, {\hat \Omega}^r$, ${\hat \sigma}^2 = {\hat \Omega}^2 =
1$). At this distance from the rotation axis the tangential
rotational velocity becomes equal to the velocity of light. This is
the {\it horizon problem} of the rotating disk (the horizon is often
named the {\it light cylinder}). Let us remark that even if in the
existing theory of rotating relativistic stars \cite{10} one uses
differential rotations, notwithstanding that in the study of the
magnetosphere of pulsars often the notion of light cylinder is still
used.

\bigskip

The search of admissible 3+1 splittings with non-Euclidean 3-spaces
is much more difficult. The simplest case is the following
parametrization of the embeddings (\ref{1.4}) in terms of Lorentz
matrices $\Lambda^A{}_B(\tau, \sigma)\, \rightarrow_{\sigma
\rightarrow \infty}\, \delta^A_B$ \footnote{It corresponds to the
{\it locality hypothesis} of Ref.\cite{6}, according to which at
each instant of time the detectors of an accelerated observer give
the same indications as the detectors of the instantaneously
comoving inertial observer.} with $\Lambda^A{}_B(\tau, 0)$ finite.
The Lorentz matrix is written in the form $\Lambda = {\cal B}\,
{\cal R}$ as the product of a boost ${\cal B}(\tau, \sigma)$ and a
rotation ${\cal R}(\tau, \sigma)$ like the one in Eq.(\ref{1.7})
(${\cal R}^{\tau}{}_{\tau} = 1$, ${\cal R}^{\tau}{}_r = 0$, ${\cal
R}^r{}_s = R^r{}_s$). The components of the boost are ${\cal
B}^{\tau}{}_{\tau}(\tau, \sigma) = \gamma(\tau, \sigma) = 1/ \sqrt{1
- {\vec \beta}^2(\tau, \sigma)}$, ${\cal B}^{\tau}{}_r(\tau, \sigma)
= \gamma(\tau, \sigma)\, \beta_r(\tau, \sigma)$, ${\cal
R}^r{}_s(\tau, \sigma) = \delta^r_s + {{\gamma\, \beta^r\,
\beta_s}\over {1 + \gamma}}(\tau, \sigma)$, with $\beta^r(\tau,
\sigma) = G(\sigma)\, \beta^r(\tau)$, where $\beta^r(\tau)$ is
defined by the 4-velocity of the observer $u^{\mu}(\tau) =
\epsilon^{\mu}_A\, \beta^A(\tau)/ \sqrt{1 - {\vec \beta}^2(\tau)}$,
$\beta^A(\tau) = (1; \beta^r(\tau))$. The M$\o$ller conditions are
restrictions on $G(\sigma)\, \rightarrow_{\sigma \rightarrow
\infty}\, 0$ with $G(0)$ finite, whose explicit form is still under
investigation.

\bigskip

The embedding (\ref{1.7}) has been studied in details in
Ref.\cite{11} for the development of quantum mechanics in
non-inertial frames.
\medskip

See the second paper of Ref.\cite{7} for the description of the
electro-magnetic field and of phenomena like the Sagnac effect and
the Faraday rotation in this framework for non-inertial frames.

\bigskip

The previous approach based on the 3+1 point of view has allowed a
complete reformulation of relativistic particle mechanics  in SR
\cite{7,12,13}. By means of {\it parametrized Minkowski theories}
\cite{12}, \cite{7}, one can get the description of arbitrary
isolated systems (particles, strings, fluids, fields) admitting a
Lagrangian formulation in arbitrary non-inertial frames
\footnote{See Ref.\cite{5} for the definition of parametrized
Galilei theories in NR mechanics.}. To get it the Lagrangian is
coupled to an external gravitational field and then the
gravitational 4-metric is replaced with the 4-metric
${}^3g_{AB}(\tau, \sigma^r)$, a functional of the embedding
$z^{\mu}(\tau, \sigma^r)$, induced by an admissible 3+1 splitting of
Minkowski space-time. The new Lagrangian, a function of the matter
and of the embedding, is invariant under the frame-preserving
diffeomorphisms of Ref.\cite{14} \footnote{This is the only paper
known to us where here is an attempt to formulate a theory of
non-inertial frames in SR.}. This kind of general covariance implies
that the embeddings are {\it gauge variables}, so that the
transition among non-inertial frames is described as a {\it gauge
transformation}: only the appearances change, not the physics.
\medskip

This framework allows us to define the {\it inertial and
non-inertial rest frames} of the isolated systems, where to develop
the rest-frame instant form of the dynamics and to build the
explicit form of the Lorentz boosts for interacting systems. While
the inertial rest frames have their Euclidean 3-spaces defined as
space-like 3-manifolds of Minkowski space-time orthogonal to the
conserved 4-momentum of the isolated system, the non-inertial rest
frames are admissible non-inertial frames whose 3-spaces tend to
those of some inertial rest frame at spatial infinity, where the
3-space becomes orthogonal to the conserved 4-momentum.\medskip

This makes possible to study the problem of the relativistic center
of mass with the associated external and internal (i.e. inside the
3-space) realizations of the Poincar\'e algebra\cite{15},
relativistic bound states \cite{16,17,18}, relativistic kinetic
theory and relativistic micro-canonical ensemble \cite{19} and
various other systems \cite{20,21}. Moreover a Wigner-covariant
relativistic quantum mechanics \cite{22}, with a solution of all the
known problems introduced by SR, has been developed after some
preliminary work done in Ref.\cite{11}. This allows the beginning of
the study of relativistic entanglement taking into account all the
consequences of the Lorentz signature of Minkowski space-time. As
shown in Ref.\cite{22} in SR the relativistic center of mass is  a
non-local non-measurable quantity: only relative variables have an
operational meaning and this implies a spatial non-separability,
i.e. some form of weak relationism in which all the objects know
each other differently from the non-relativistic case where the
center of a mass is a measurable quantity.

\medskip

See Ref.\cite{23} for an extended review of this approach both in SR
and in GR. In the next Section there will be a sketch of the known
results in GR.

\section{Non-Inertial Frames in General Relativity}

In GR global inertial frames are forbidden by the equivalence
principle. Therefore gravitational physics has to be described in
non-inertial frames.\medskip

While in SR Minkowski space-time is an absolute notion, unifying the
absolute notions of time and 3-space of the NR Galilei space-time,
in Einstein GR also the space-time is a dynamical object \cite{24}
and the gravitational field is described by the metric structure of
the space-time, namely by the ten dynamical fields
${}^4g_{\mu\nu}(x)$ ($x^{\mu}$ are world 4-coordinates) satisfying
Einstein equations.
\medskip

The ten dynamical fields ${}^4g_{\mu\nu}(x)$ are not only a
(pre)potential for the gravitational field (like the
electro-magnetic and Yang-Mills fields are the potentials for
electro-magnetic and non-Abelian forces) but also determines the
{\it chrono-geometrical structure of space-time} through the line
element $ds^2 = {}^4g_{\mu\nu}\, dx^{\mu}\, dx^{\nu}$. Therefore the
4-metric teaches relativistic causality to the other fields: it says
to massless particles like photons and gluons which are the allowed
world-lines in each point of space-time. The ACES mission of ESA
\cite{25} will give the first precision measurement of the
gravitational red-shift of the geoid, namely of the $1/c^2$
deformation of Minkowski light-cone caused by the geo-potential.

\bigskip

The metrology-oriented solution of the problem of clock
synchronization used in SR can be extended to GR if Einstein
space-times are restricted to the class of globally hyperbolic,
topologically trivial, asymptotically Minkowskian space-times
without super-translations \footnote{At this preliminary level these
space-times must also be without Killing symmetries, because,
otherwise, at the Hamiltonian level one should introduce complicated
sets of extra Dirac constraints for each existing Killing vector.}.

\medskip

As shown in the first paper of Ref.\cite{26}, in the chosen class of
space-times the 4-metric ${}^4g_{\mu\nu}(x)$ tends in a suitable way
to the flat Minkowski 4-metric ${}^4\eta_{\mu\nu}$ at spatial
infinity and the ten {\it strong} asymptotic ADM Poincar\'e
generators $P^A_{ADM}$, $J^{AB}_{ADM}$ (they are fluxes through a
2-surface at spatial infinity) are well defined functionals of the
4-metric fixed by the boundary conditions at spatial infinity.
\medskip

These properties do not hold in generic asymptotically flat
space-times, because they have the SPI group of asymptotic
symmetries (direction-dependent asymptotic Killing symmetries)
\cite{27} and this is an obstruction to the existence of asymptotic
Lorentz generators for the gravitational field \cite{28}. However if
one restricts the class of space-times to those {\it not containing
super-translations} \cite{29}, then the SPI group reduces to the
asymptotic ADM Poincar\'e group \cite{30}: these space-times are
{\it asymptotically Minkowskian}, they contain an asymptotic
Minkowski 4-metric (to be used as an {\it asymptotic background} at
spatial infinity in the linearization of the theory) and they have
asymptotic inertial observers at spatial infinity whose spatial axes
may be identified by means of the fixed stars of star catalogues
\footnote{The fixed stars can be considered as an empirical
definition of spatial infinity of the observable universe.}.
Moreover, in the limit of vanishing Newton constant ($ G = 0$) the
asymptotic ADM Poincar\'e generators become the generators of the
special relativistic Poincar\'e group describing the matter present
in the space-time. This is an important condition for the inclusion
into GR of the classical version of the standard model of particle
physics, whose properties are all connected with the representations
of this group in the inertial frames of Minkowski space-time. In
absence of matter a sub-class of these space-times is the
(singularity-free) family of Chrstodoulou-Klainermann solutions of
Einstein equations \cite{31} (they are near to Minkowski space-time
in a norm sense and contain gravitational waves).

\bigskip

In the first paper of Ref.\cite{26} it is also shown that the
boundary conditions on the 4-metric required by the absence of
super-translations imply that the only admissible 3+1 splittings of
space-time (i.e. the allowed global non-inertial frames) are the
{\it non-inertial rest frames}:  their 3-spaces are asymptotically
orthogonal to the weak ADM 4-momentum. Therefore one gets ${\hat
P}^r_{ADM} \approx 0$ as the rest-frame condition of the 3-universe
with a mass and a rest spin fixed by the boundary conditions. Like
in SR the 3-universe can be visualized as a decoupled non-covariant
(non-measurable) external relativistic center of mass plus an
internal non-inertial rest-frame 3-space containing only relative
variables (see the first paper in Ref.\cite{32}).

\bigskip

In these space-times one can define global non-inertial frames by
using the same admissible 3+1 splittings, centered on a time-like
observer, and the observer-dependent radar 4-coordinates $\sigma^A =
(\tau; \sigma^r)$ employed in SR. This  will allow to separate the
{\it inertial} (gauge) degrees of freedom of the gravitational field
(playing the role of inertial potentials) from the dynamical {\it
tidal} ones at the Hamiltonian level.

\medskip

In GR the dynamical fields are the components ${}^4g_{\mu\nu}(x)$ of
the 4-metric and not the  embeddings $x^{\mu} = z^{\mu}(\tau,
\sigma^r)$ defining the admissible 3+1 splittings of space-time like
in  the parametrized Minkowski theories of SR. Now the gradients
$z^{\mu}_A(\tau, \sigma^r)$ of the embeddings give the transition
coefficients from radar to world 4-coordinates, so that the
components ${}^4g_{AB}(\tau, \sigma^r) = z^{\mu}_A(\tau, \sigma^r)\,
z^{\nu}_B(\tau, \sigma^r)\, {}^4g_{\mu\nu}(z(\tau, \sigma^r))$ of
the 4-metric will be the dynamical fields in the ADM action. Like in
SR the 4-vectors $z^{\mu}_A(\tau, \sigma^r)$, tangent to the
3-spaces $\Sigma_{\tau}$, are used to define the unit normal
$l^{\mu}(\tau, \sigma^r) = z^{\mu}_A(\tau, \sigma^r)\, l^A(\tau,
\sigma^r)$ to $\Sigma_{\tau}$, while the 4-vector
$z^{\mu}_{\tau}(\tau, \sigma^r)$ has the lapse function as component
along the unit normal and the shift functions as components along
the tangent vectors.

\medskip

Since the world-line of the time-like observer can be chosen as the
origin of a set of the spatial world coordinates, i.e.
$x^{\mu}(\tau) = (x^o(\tau); 0)$, it turns out that with this choice
the space-like surfaces of constant coordinate time $x^o(\tau) =
const.$ coincide with the dynamical instantaneous 3-spaces
$\Sigma_{\tau}$ with $\tau = const.$. By using asymptotic flat
tetrads $\epsilon^{\mu}_A = \delta^{\mu}_o\, \delta^{\tau}_A +
\delta^{\mu}_i\, \delta^i_A$ (with $\epsilon^A_{\mu}$ denoting the
inverse flat cotetrads) and by choosing a coordinate world time
$x^o(\tau) = x^o_o + \epsilon^o_{\tau}\, \tau = x^o_o + \tau$, one
gets the following preferred embedding corresponding to these given
world 4-coordinates

\beq
 x^{\mu} = z^{\mu}(\tau, \sigma^r) = x^{\mu}(\tau) +
 \epsilon^{\mu}_r\, \sigma^r = \delta^{\mu}_o\, x^o_o +
 \epsilon^{\mu}_A\, \sigma^A.
 \label{2.1}
 \eeq

\noindent This choice implies $z^{\mu}_A(\tau, \sigma^r) =
\epsilon^{\mu}_A$ and ${}^4g_{\mu\nu}(x = z(\tau, \sigma^r)) =
\epsilon^A_{\mu}\, \epsilon_{\nu}^B\, {}^4g_{AB}(\tau, \sigma^r)$.

\bigskip

As shown in Ref.\cite{24}, the dynamical nature of space-time
implies that each solution (i.e. an Einstein 4-geometry) of
Einstein's equations (or of the associated ADM Hamilton equations)
dynamically selects a preferred 3+1 splitting of the space-time,
namely in GR the instantaneous 3-spaces  are dynamically determined
in the chosen world coordinate system, modulo the choice of the
3-coordinates in the 3-space and modulo the trace of the extrinsic
curvature of the 3-space as a space-like sub-manifold of the
space-time. Eq.(\ref{2.1}) can be used to describe this 3+1
splitting and then by means of 4-diffeomorphisms the solution can be
written in an arbitrary world 4-coordinate system in general not
adapted to the dynamical 3+1 splitting. This gives rise to the
4-geometry corresponding to the given solution.

\bigskip

To define the canonical formalism the Einstein-Hilbert action for
metric gravity (depending on the second derivative of the metric)
must be replaced with the ADM action (the two actions differ for a
surface tern at spatial infinity). As shown in the first paper of
Refs.\cite{26}, the Legendre transform and the definition of a
consistent canonical Hamiltonian require the introduction of the
DeWitt surface term at spatial infinity: the final canonical
Hamiltonian turns out to be the {\it strong} ADM energy (a flux
through a 2-surface at spatial infinity), which is equal to the {\it
weak} ADM energy (expressed as a volume integral over the 3-space)
plus constraints.
\medskip

Therefore there is not a frozen picture like in the "spatially
compact space-times without boundaries" used in loop quantum gravity
\footnote{In these space-times the canonical Hamiltonian vanishes
and the Dirac Hamiltonian is  a combination of first-class
constraints, so that it only generates Hamiltonian gauge
transformations. In the reduced phase space, quotient with respect
the Hamilonian gauge group, the reduced Hamiltonian is zero and one
has a {\it frozen picture} of dynamics. This class of space-times
fits well with Machian ideas (no boundary conditions) and with
interpretations in which there is no physical time like the one in
Ref.\cite{33}. However, it is not clear how to include in this
framework the standard model of particle physics.}, but an evolution
generated by a Dirac Hamiltonian equal to the weak ADM energy plus a
linear combination of the first class constraints. Also the other
strong ADM Poincar\'e generators are replaced by their weakly
equivalent weak form ${\hat P}^A_{ADM}$, ${\hat
J}^{AB}_{ADM}$.\bigskip

To take into account the fermion fields present in the standard
particle model one must extend ADM gravity  to ADM tetrad gravity .
Since our class of space-times admits orthonormal tetrads and a
spinor structure \cite{34}, the extension can be done by simply
replacing the 4-metric in the ADM action with its expression in
terms of tetrad fields, considered as the basic 16 configurational
variables substituting the 10 metric fields. This can be achieved by
decomposing the 4-metric on cotetrad fields (by convention a sum on
repeated indices is assumed)

\beq
 {}^4g_{AB}(\tau, \sigma^r) = E_A^{(\alpha)}(\tau, \sigma^r)\,
 {}^4\eta_{(\alpha)(\beta)}\, E^{(\beta)}_B(\tau, \sigma^r),
 \label{2.2}
 \eeq

\noindent by putting this expression into the ADM action and by
considering the resulting action, a functional of the 16 fields
$E^{(\alpha)}_A(\tau, \sigma^r)$, as the action for ADM tetrad
gravity. In Eq.(\ref{2.2}) $(\alpha)$ are flat indices and the
cotetrad fields $E^{(\alpha)}_A$ are the inverse of the tetrad
fields $E^A_{(\alpha)}$, which are connected to the world tetrad
fields by $E^{\mu}_{(\alpha)}(x) = z^{\mu}_A(\tau, \sigma^r)\,
E^A_{(\alpha)}(z(\tau, \sigma^r))$ by the embedding of
Eq.(\ref{2.1}).

\medskip

This leads to an interpretation of gravity based on a congruence of
time-like observers endowed with orthonormal tetrads: in each point
of space-time the time-like axis is the  unit 4-velocity of the
observer, while the spatial axes are a (gauge) convention for
observer's gyroscopes. This framework was developed in the second
and third paper of  Refs.\cite{26}.
\medskip

Even if the action of ADM tetrad gravity depends upon 16 fields, the
counting of the physical degrees of freedom of the gravitational
field does not change, because this action is invariant not only
under the group of 4-difeomorphisms but also under the O(3,1) gauge
group of the Newman-Penrose approach \cite{35} (the extra gauge
freedom acting on the tetrads in the tangent space of each point of
space-time).

\medskip

The cotetrads $E^{(\alpha)}_A(\tau, \sigma^r)$ are the new
configuration variables. They are connected to cotetrads
$\eo^{(\alpha )}_A(\tau, \sigma^r)$ adapted to the 3+1 splitting of
space-time, namely such that the inverse adapted time-like tetrad
$\eo_{(o)}^A(\tau, \sigma^r)$ is the unit normal to the 3-space
$\Sigma_{\tau}$, by a standard Wigner boosts for time-like
Poincar\'e orbits with parameters $\varphi_{(a)}(\tau, \sigma^r)$,
$a=1,2,3$

\bea
 E_A^{\alpha)} &=& L^{(\alpha)}{}_{(\beta)}( \varphi_{(a)})\,
 {\buildrel o\over E}_A^{(\beta)}, \qquad {}^4g_{AB} = \eo^{(\alpha
 )}_A\, {}^4\eta_{(\alpha )(\beta )}\, \eo^{(\beta )}_B,\nonumber \\
 &&{}\nonumber \\
 L^{(\alpha )}{}_{(\beta )}(\varphi_{(a)}) &{\buildrel {def}\over
 =}& L^{(\alpha )}{}_{(\beta )}(V(z(\sigma ));\,\, {\buildrel \circ
 \over V}) = \delta^{(\alpha )}_{(\beta )} + 2 \sgn\, V^{(\alpha
 )}(z(\sigma ))\, {\buildrel \circ \over V}_{(\beta )} -\nonumber \\
 &-&\sgn\, {{(V^{(\alpha )}(z(\sigma )) + {\buildrel \circ \over
 V}^{(\alpha )})\, (V_{(\beta )}(z(\sigma )) + {\buildrel \circ \over
 V}_{(\beta )})}\over {1 + V^{(o)}(z(\sigma ))}}.
 \label{2.3}
 \eea

\noindent In each tangent plane to a point of $\Sigma_{\tau}$ this
point-dependent standard Wigner boost sends the unit future-pointing
time-like vector ${\buildrel o\over V}^{(\alpha )} = (1; 0)$ into
the unit time-like vector $V^{(\alpha )} = {}^4E^{(\alpha )}_A\, l^A
= \Big(\sqrt{1 + \sum_a\, \varphi^2_{(a)}}; \varphi^{(a)} = - \sgn\,
\varphi_{(a)}\Big)$. As a consequence, the flat indices $(a)$ of the
adapted tetrads and cotetrads and of the triads and cotriads on
$\Sigma_{\tau}$ transform as Wigner spin-1 indices under
point-dependent SO(3) Wigner rotations $R_{(a)(b)}(V(z(\sigma
));\,\, \Lambda (z(\sigma ))\, )$ associated with Lorentz
transformations $\Lambda^{(\alpha )}{}_{(\beta )}(z)$ in the tangent
plane to the space-time in the given point of $\Sigma_{\tau}$.
Instead the index $(o)$ of the adapted tetrads and cotetrads is a
local Lorentz scalar index.

\bigskip

The adapted tetrads and cotetrads   have the expression

\bea
 \eo^A_{(o)} &=& {1\over {1 + n}}\, (1; - \sum_a\, n_{(a)}\,
 {}^3e^r_{(a)}) = l^A,\qquad \eo^A_{(a)} = (0; {}^3e^r_{(a)}), \nonumber \\
 &&{}\nonumber  \\
 \eo^{(o)}_A &=& (1 + n)\, (1; \vec 0) = \sgn\, l_A,\qquad \eo^{(a)}_A
= (n_{(a)}; {}^3e_{(a)r}),
 \label{2.4}
 \eea

\noindent where ${}^3e^r_{(a)}$ and ${}^3e_{(a)r}$ are triads and
cotriads on $\Sigma_{\tau}$ and $n_{(a)} = n_r\, {}^3e^r_{(a)} =
n^r\, {}^3e_{(a)r}$ \footnote{Since one uses the positive-definite
3-metric $\delta_{(a)(b)} $, one will use only lower flat spatial
indices. Therefore for the cotriads one uses the notation
${}^3e^{(a)}_r\,\, {\buildrel {def}\over =}\, {}^3e_{(a)r}$ with
$\delta_{(a)(b)} = {}^3e^r_{(a)}\, {}^3e_{(b)r}$.} are adapted shift
functions. In Eqs.(\ref{2.4}) $N(\tau, \vec \sigma) = 1 + n(\tau,
\vec \sigma) > 0$, with $n(\tau ,\vec \sigma)$ vanishing at spatial
infinity (absence of super-translations), so that $N(\tau, \vec
\sigma)\, d\tau$ is positive from $\Sigma_{\tau}$ to $\Sigma_{\tau +
d\tau}$, is the lapse function; $N^r(\tau, \vec \sigma) = n^r(\tau,
\vec \sigma)$, vanishing at spatial infinity (absence of
super-translations), are the shift functions.

\bigskip

The adapted tetrads $\eo^A_{(a)}$ are defined modulo SO(3) rotations
$\eo^A_{(a)} = \sum_b\, R_{(a)(b)}(\alpha_{(e)})\, {}^4{\buildrel
\circ \over {\bar E}}^A_{(b)}$, ${}^3e^r_{(a)} = \sum_b\,
R_{(a)(b)}(\alpha_{(e)})\, {}^3{\bar e}^r_{(b)}$, where
$\alpha_{(a)}(\tau ,\vec \sigma )$ are three point-dependent Euler
angles. After having chosen an arbitrary point-dependent origin
$\alpha_{(a)}(\tau ,\vec \sigma ) = 0$, one arrives at the following
adapted tetrads and cotetrads [${\bar n}_{(a)} = \sum_b\, n_{(b)}\,
R_{(b)(a)}(\alpha_{(e)})\,$, $\sum_a\, n_{(a)}\, {}^3e^r_{(a)} =
\sum_a\, {\bar n}_{(a)}\,
 {}^3{\bar e}^r_{(a)}$]

\bea
 {}^4{\buildrel \circ \over {\bar E}}^A_{(o)}
 &=& \eo^A_{(o)} = {1\over {1 + n}}\, (1; - \sum_a\, {\bar n}_{(a)}\,
 {}^3{\bar e}^r_{(a)}) = l^A,\qquad {}^4{\buildrel \circ \over
 {\bar E}}^A_{(a)} = (0; {}^3{\bar e}^r_{(a)}), \nonumber \\
 &&{}\nonumber  \\
 {}^4{\buildrel \circ \over {\bar E}}^{(o)}_A
 &=& \eo^{(o)}_A = (1 + n)\, (1; \vec 0) = \sgn\, l_A,\qquad
 {}^4{\buildrel \circ \over {\bar E}}^{(a)}_A
= ({\bar n}_{(a)}; {}^3{\bar e}_{(a)r}),
 \label{2.5}
 \eea

\noindent which one will use as a reference standard.\medskip

The expression for the general tetrad

\bea
 {}^4E^A_{(\alpha )} &=& \eo^A_{(\beta )}\, L^{(\beta )}{}_{(\alpha
 )}(\varphi_{(a)}) = {}^4{\buildrel \circ \over {\bar E}}^A_{(o)}\,
 L^{(o)}{}_{(\alpha )}(\varphi_{(c)}) +\nonumber \\
 &+& \sum_{ab}\, {}^4{\buildrel \circ \over
 {\bar E}}^A_{(b)}\, R^T_{(b)(a)}(\alpha_{(c)})\,
 L^{(a)}{}_{(\alpha )}(\varphi_{(c)}),
 \label{2.6}
 \eea
\medskip

\noindent shows that every point-dependent Lorentz transformation
 $\Lambda$ in the tangent planes may be parametrized with the
 (Wigner) boost parameters $\varphi_{(a)}$ and the Euler angles
 $\alpha_{(a)}$, being the product $\Lambda = R\, L$ of a rotation
 and a boost.

\bigskip

The future-oriented unit normal to $\Sigma_{\tau}$ and the projector
on $\Sigma_{\tau}$ are $l_A = \sgn\, (1 + n)\, \Big(1;\, 0\Big)$,
${}^4g^{AB}\, l_A\, l_B = \sgn $, $l^A = \sgn\, (1 + n)\,
{}^4g^{A\tau} = {1\over {1 + n}}\, \Big(1;\, - n^r\Big) = {1\over {1
+ n}}\, \Big(1;\, - \sum_a\, {\bar n}_{(a)}\, {}^3{\bar
e}_{(a)}^r\Big)$, ${}^3h^B_A = \delta^B_A - \sgn\, l_A\, l^B$.

\bigskip

The 4-metric has the following expression

 \bea
 {}^4g_{\tau\tau} &=& \sgn\, [(1 + n)^2 - {}^3g^{rs}\, n_r\,
 n_s] = \sgn\, [(1 + n)^2 - \sum_a\, {\bar n}^2_{(a)}],\nonumber \\
 {}^4g_{\tau r} &=& - \sgn\, n_r = -\sgn\, \sum_a\, {\bar n}_{(a)}\,
 {}^3{\bar e}_{(a)r},\nonumber \\
  {}^4g_{rs} &=& -\sgn\, {}^3g_{rs} = - \sgn\, \sum_a\, {}^3e_{(a)r}\, {}^3e_{(a)s}
  = - \sgn\, \sum_a\, {}^3{\bar e}_{(a)r}\, {}^3{\bar e}_{(a)s},\nonumber \\
 &&{}\nonumber \\
 {}^4g^{\tau\tau} &=& {{\sgn}\over {(1 + n)^2}},\qquad
  {}^4g^{\tau r} = -\sgn\, {{n^r}\over {(1 + n)^2}} = -\sgn\, {{\sum_a\, {}^3{\bar e}^r_{(a)}\,
 {\bar n}_{(a)}}\over {(1 + n)^2}},\nonumber \\
 {}^4g^{rs} &=& -\sgn\, ({}^3g^{rs} - {{n^r\, n^s}\over
 {(1 + n)^2}}) = -\sgn\, \sum_{ab}\, {}^3{\bar e}^r_{(a)}\, {}^3{\bar e}^s_{(b)}\, (\delta_{(a)(b)} -
 {{{\bar n}_{(a)}\, {\bar n}_{(b)}}\over {(1 + n)^2}}),\nonumber \\
 &&{}\nonumber \\
 &&\sqrt{- g } = \sqrt{|{}^4g|} = {{\sqrt{{}^3g}}\over {\sqrt{\sgn\,
 {}^4g^{\tau\tau}}}} = \sqrt{\gamma}\, (1 + n) = {}^3e\, (1 +
 n),\nonumber \\
 && {}^3g = \gamma = ({}^3e)^2,\quad {}^3e = det\, {}^3e_{(a)r}.
 \label{2.7}
 \eea

\bigskip

The 3-metric ${}^3g_{rs}$ has signature $(+++)$, so that one may put
all the flat 3-indices {\it down}. One has ${}^3g^{ru}\, {}^3g_{us}
= \delta^r_s$.

\bigskip

After having introduced the kinematical framework for the
description of non-inertial frames in GR, we must study the
dynamical aspects of the gravitational field to understand which
variables are dynamically determined and which are the inertial
effects hidden in the general covariance of the theory. Since at the
Lagrangian level it is not possible to identify which components of
the 4-metric tensor are connected with the gauge freedom in the
choice of the 4-coordinates and which ones describe the dynamical
degrees of freedom of the gravitational field, one must restrict
himself to the class of globally hyperbolic, asymptotically flat
space-times allowing a Hamiltonian description starting from the
description of Einstein GR in terms of the ADM action \cite{36}
instead than in terms of the Einstein-Hilbert one. In canonical ADM
gravity one can use Dirac theory of constraints \cite{37} to
describe the Hamiltonian gauge group, whose generators are the
first-class constraints of the model. The basic tool of this
approach is the possibility to find so-called Shanmugadhasan
canonical transformations \cite{38}, which identify special
canonical bases adapted to the first-class constraints (and also to
the second-class ones when present). In these special canonical
bases the vanishing of certain momenta (or of certain
configurational coordinates) corresponds to the vanishing of well
defined Abelianized combinations of the first-class constraints
(Abelianized because the new constraints have exactly zero Poisson
brackets even if the original constraints were not in strong
involution). As a consequence, the variables conjugate to these
Abelianized constraints are inertial Hamiltonian gauge variables
describing the Hamiltonian gauge freedom.

\bigskip

Starting from the ADM action for tetrad gravity one defines the
Hamiltonian formalism in a phase space containing 16 configurational
field variables and 16 conjugate moments. One identifies the 14
first-class constraints of the system and one finds that the
canonical Hamiltonian is the weak ADM energy (it is given as a
volume integral over 3-space). The existence of these 14 first-class
constraints implies that 14 components of the tetrads (or of the
conjugate momenta) are Hamiltonian gauge variables describing the
{\it inertial} aspects of the gravitational field (6 of these
inertial variables describe the extra gauge freedom in the choice of
the tetrads and in their transport along world-lines). Therefore
there are only 2+2 degrees of freedom for the description of the
{\it tidal} dynamical aspects of the gravitational field. The
asymptotic ADM Poincar\'e generators can be evaluated explicitly.
Till now the type of matter studied in this framework \cite{32}
consists of the electro-magnetic field and of N charged scalar
particles, whose signs of the energy and electric charges are
Grassmann-valued to regularize both the gravitational and
electro-magnetic self-energies (it is both a ultraviolet and an
infrared regularization),
\bigskip

The remaining 2+2 conjugate variables describe the dynamical tidal
degrees of freedom of the gravitational field (the two polarizations
of gravitational waves in the linearized theory). If one would be
able to include all the constraints in the Shanmugadhasan canonical
basis, these 2+2 variables would be the {\it Dirac observables} of
the gravitational field, invariant under the Hamiltonian gauge
transformations. However such Dirac observables are not known: one
only has statements bout their existence \cite{39}. Moreover, in
general they are not 4-scalar observables. The problem of the
connection between the 4-diffeomorphism group and the Hamiltonian
gauge group was studied in Ref.\cite{40} by means of the inverse
Legendre transformation and of the notion of dynamical symmetry. The
conclusion is that on the space of solutions of Einstein equations
there is an overlap of the two types of observables: there should
exists special Shanmugadhasan canonical bases in which the 2+2 Dirac
observables become 4-scalars when restricted to the space of
solutions of the Einstein equations. In any case the identification
of the inertial gauge components of the 4-metric is what is needed
to make a fixation of 4-coordinates as required by relativistic
metrology.

\bigskip

It can be shown that there is a Shanmugadhasan canonical
transformation \cite{41} (implementing the so-called York map
\cite{42} and diagonalizing the York-Lichnerowics approach
\cite{43}) to a so-called York canonical basis adapted to 10 of the
14 first-class constraints. Only the super-Hamiltonian and
super-momentum constraints, whose general solution is not known, are
not included in the basis, but it is clarified which variables are
to be determined by their solution, namely the 3-volume element (the
determinant of the 3-metric) of the 3-space $\Sigma_{\tau}$ and the
three momenta conjugated to the 3-coordinates on $\Sigma_{\tau}$.
The 14 inertial gauge variables turn out to be: a) the six
configurational variables $\varphi_{(a)}$ and $\alpha_{(a)}$ of the
tetrads describing their O(3,1) gauge freedom; b) the lapse and
shift functions; c) the 3-coordinates on the 3-space (their fixation
implies the determination of the shift functions); d) the York time
\cite{44} ${}^3K$, i.e. the trace of the extrinsic curvature of the
3-spaces as 3-manifolds embedded into the space-time (its fixation
implies the determination of the lapse function). It is the only
gauge variable which is a momentum in the York canonical basis
\footnote{Instead in Yang-Mills theory all the gauge variables are
configurational.}: this is due to the Lorentz signature of
space-time, because the York time and three other inertial gauge
variables can be used as 4-coordinates of the space-time (see
Ref.\cite{24} for this topic and for its relevance in the solution
of the hole argument). In this way  an identification of the
inertial gauge variables to be fixed to get a 4-coordinate system in
relativistic metrology was found. While in SR all the components of
the tetrads and their conjugate momenta are inertial gauge
variables, in GR the two eigenvalues of the 3-metric with
determinant one and their conjugate momenta describe the physical
tidal degrees of freedom of the gravitational field. In the first
paper of Ref.\cite{32} there is the expression of the Hamilton
equations for all the variables of the York canonical basis.

\bigskip

An important remark is that in the framework of the York canonical
basis the natural family of gauges is not the harmonic one, but the
family of 3-orthogonal Schwinger time gauges in which the 3-metric
in the 3-spaces is diagonal.

\bigskip

Both in SR and GR an admissible 3+1 splitting of space-time has two
associated congruences of time-like observers \cite{7},
geometrically defined and not to be confused with the congruence of
the world-lines of fluid elements, when relativistic fluids are
added as matter in GR \cite{45,46,47}. One of the two congruences,
with zero vorticity, is the congruence of the Eulerian observers,
whose 4-velocity field is the field of unit normals to the 3-spaces.
This congruence allows us to re-express the non-vanishing momenta of
the York canonical basis in terms of the expansion ($\theta = -
{}^3K$) and of the shear of the Eulerian observers. This allows us
to compare the Hamilton equations of ADM canonical gravity with the
usual first-order non-Hamiltonian ADM equations deducible from
Einstein equations given a 3+1 splitting of space-time but without
using the Hamiltonian formalism. As a consequence, one can extend
our Hamiltonian identification of the inertial and tidal variables
of the gravitational field to the Lagrangian framework and use it in
the cosmological (conformally asymptotically flat) space-times: in
them it is not possible to formulate the Hamiltonian formalism but
the standard ADM equations are well defined. The time inertial gauge
variable needed for relativistic metrology is now the expansion of
the Eulerian observers of the given 3+1 splitting of the globally
hyperbolic cosmological space-time.

\section{Conclusion}

In conclusion we now have a framework for non-inertial frames in GR
and an identification of the inertial gauge variables in
asymptotically Minkowskian and also cosmological space-times.

\medskip

See Refs.\cite{23,32} for the possibility that dark matter  is only
a relativistic inertial effect induced by the inertial gauge
variable ${}^3K$ (the York time): a suitable choice of the 3-space
in PM Celestial Reference Frame could simulate the effects explained
with dark matter.

\medskip

Moreover in Ref.\cite{23}, at a preliminary level, it is also shown
that the York time is connected also with dark energy in
cosmological space-times \cite{4}. In the standard FWR space-times
the Killing symmetries connected with homogeneity and isotropy imply
($\tau$ is the cosmic time, $a(\tau)$ the scale factor) that the
York time is no more a gauge variable but coincides with the Hubble
constant: ${}^3K (\tau) = - {{\dot a(\tau)}\over {a(\tau)}} = -
H(\tau)$. However at the first order in cosmological perturbations
(see Ref.\cite{48} for a review) one has ${}^3K = - H + {}^3K_{(1)}$
with ${}^3K_{(1)}$ being again an inertial gauge variable to be
fixed with a metrological convention. Therefore the York time has a
central role also in cosmology and one needs to know the dependence
on it of the main quantities, like the red-shift and the luminosity
distance from supernovae \cite{49}, which  require the introduction
of the notion of dark energy to explain the 3-universe and its
accelerated expansion in the framework of the standard $\Lambda$CDM
cosmological model.\medskip

In particular it will be important to study inhomogeneous
space-times without Killing symmetries like the Szekeres ones
\cite{50}, where the York time remains an arbitrary inertial gauge
variable, to see whether it is possible to find a 3-orthogonal gauge
in them (at least in a PM approximation) in which the convention on
the inertial gauge variable York time allows one to eliminate both
dark matter and dark energy through the choice of a 4-coordinate
system  in a consistent PM reformulation of ICRS and simultaneously
to save the main good properties of the standard $\Lambda$CDM
cosmological model due to the inertial and dynamical properties of
the space-time.

\vfill\eject

\end{document}